# Ionization potential depression model with the influence of neighboring ions for warm/hot and dense plasma

Chensheng Wu[1,2,a], Jiao Sun[2,a], Qinghe Song[1], Chunhua Zeng[1], Xiang Gao[2,*] , and Jun Yan[2,†]

[1] *Faculty of Science, Kunming University of Science and Technology, Kunming 650500, China and*

[2] *Institute for Applied Physics and Computational Mathematics, Beijing 100088, China*

[a]*These authors contributed equally to this work.*

* *gao_xiang@iapcm.ac.cn*

† *yan_jun@iapcm.ac.cn*

**Abstract**

For warm/hot and dense plasma (WDP), ionization potential depression (IPD) plays a crucial role in determining the ionization balance and understanding the resulting microscopic plasma properties. However, the applicability of the widely used IPD models is currently limited under WDP conditions, where the influence of neighboring ions on ionization potential depression (IPD) becomes nonnegligible. Neighboring ions can directly influence the screening potential around the target ion, which then changes the ionization potential. Furthermore, similar to solid-state systems, outer atomic orbitals expand into continuous energy bands due to the existence of neighboring ions, and electrons in these continuous bands can travel from target ion into neighboring ions and become delocalized. As a result, even for their total energy $E<0$, electrons excited into these continuous bands can be considered ionized, and the ionization conditions differ from those in isolated situations. In our previous work with an atomic-state-dependent screening model, we included the influence of temporarily recombined electron distributions due to inelastic collision processes between plasma electrons and ions, and evident contributions from these electrons to the screening potential were found under WDP conditions. We now further incorporate the direct contributions of neighboring ions to both screening potentials and ionization conditions. This extended framework reveals that the contribution from neighboring ions substantially influences IPD in WDP. The developed model demonstrates good agreement with experiments for Al, Mg, and Si plasmas with a wide range of 70–700 eV temperatures and 1–3 times the solid density as well as the hollow Al ions measured in the experiment. With lower computational costs and wider application range of plasma conditions, the developed model is expected to provide a promising tool to study the ionization balance and the atomic processes as well as the related radiation and particle transport properties for a wide range of WDP.

**Keywords:** ionization potential depression; Warm and dense plasma; neighbouring ions effects; electron delocalization

# Introduction

Warm/hot and dense plasma (WDP) widely exists in stars or giant planets[1, 2] and can be created in experiments with high-power lasers and Z pinches[3-11]. For atoms embedded in a dense plasma, the atomic parameters are significantly influenced by the plasma screening due to the complicated many-body interactions with the surrounding plasma[12-21]. Ionization potential depression (IPD) is one of the most important phenomena of plasma screening. In dense plasma, IPD can significantly alter the ionization balance[22-24] and further impact several important optical and thermodynamic properties of the plasma, such as opacity and equation of state (EOS)[25, 26]. Therefore, IPD is of fundamental importance for related studies of astrophysics, inertial confinement fusion and matter under extreme conditions.

Owing to the quantum many-body correlations between electrons and ions, an accurate description of IPD in dense plasma is extremely complicated. For convenience of application, semiempirical models such as the Stewart–Pyatt (SP)[4, 5, 27] and Ecker–Kröll (EK)[6, 28] models have been proposed and widely applied in the numerical modeling and analysis of plasma since the 1960s. Recently, the development of X-ray free-electron lasers (XFELs) has made it possible to measure IPD precisely in laboratories[4-6]. However, a recent XFEL experiment in solid density Al plasma revealed that the SP model underestimates the IPD, whereas the earlier Ecker–Kröll model is consistent with the measurements[4, 5]. Moreover, in later measurements of IPD with more element species and plasma conditions, both the EK model and the SP model clearly deviated from the experimental results[6]. Owing to the discrepancy between experiments and theory, a more fundamental understanding of warm and dense plasma is urgently needed. Therefore, to address the complexity of quantum many-body interactions in plasmas, a series of sophisticated models have been developed. These models aim to capture essential plasma characteristics within specific parameter ranges while maintaining a balance between computational feasibility and physical accuracy.

Models such as finite-temperature density functional (DFT) simulations[29, 30],

atomic–solid–plasma models[31], classical molecular dynamics simulations[32, 33] and quantum statistical models[34-37] are based mainly on quantum statistical theory from an ensemble perspective, with their core focus on the dielectric function that characterizes the system response and its closely related dynamic structure factor. However, interactions in plasma are often treated through substantial approximations to manage computational complexity, and most models are restricted to local thermodynamic equilibrium (LTE) conditions, limiting their applicability to non-LTE plasma regimes. However, these models are restricted by high computational costs, and accurate treatment of ions in highly excited states remains difficult.

In contrast to ensemble-based approaches, a distinct class of plasma models employs single-atom approximations, including the Debye–Hückel (DH) model[38], two-step Hartree–Fock (HF) model[39], self-consistent ion sphere (IS) model[23, 40] and our recently established atomic-state-dependent screening model[41, 42]. In these models, detailed atomic configurations and complicated many-body correlations and dynamic atomic processes in ions can be included; thus, the influence of plasma electrons that deviate from LTE conditions can be discussed. Compared with models based on quantum statistical theory, single-atom approximation models have lower computational costs and are more flexible in application. Notably, the theoretical connection between these approaches is established through quantum statistical theory. Specifically, the Debye–Hückel model emerges as a limiting case within the random phase approximation (RPA) framework under static conditions[34], thereby bridging single-atom models with more fundamental quantum statistical treatments.

Despite their computational efficiency, single-atom approximation models exhibit limitations in WDP conditions: the influence of neighboring ions becomes non-negligible. The neighboring ions can directly influence the screening potential around the target ion, and similar to solid-state systems, outer atomic orbitals expand into continuous energy bands due to neighboring ion interactions. Electrons distributed in these continuous bands can travel from the target ion into neighboring ions and become delocalized, and electrons excited into these continuous bands can be considered ionized even if their total energy is $E < 0$, which is different from the situation of isolated ions, where electrons can be considered ionized when they are excited into states with $E>0$. In previous single-atom approximation models, the direct contributions of neighboring ions to the screening potential were not considered. By including "inner-ionization" correction in IPD, some models have attempted to consider the change in ionization conditions for ions in dense plasma[39]. However, as the delocalization of electrons in the continuous energy bands was not considered, the contribution from the changes in ionization conditions might be underestimated.

In the plasma environment, plasma electrons can temporarily recombine with ions and distribute into negative-energy states with $E<0$ through three-body recombination (TBR), dielectronic recombination (DR) and other inelastic collision processes between electrons and ions. These temporarily recombined electrons also contribute to plasma screening, which are not considered in previous models. In our recent works on atomic-state-dependent screening models[41, 42], the contributions from these temporarily recombined electrons with non-LTE distributions due to different atomic processes were considered in a unified framework. However, similar to the other previous models, the direct contributions from neighboring ions to the screening potential are not included.

In this work, we refine the atomic-state-dependent screening model by further considering the direct contribution of neighboring ions to the screening potential and the changes in ionization conditions due to neighboring ions. The developed model is successfully validated by well reproducing the IPDs in recent WDP experiments of Al, Mg and Si plasmas under wide-range conditions of 70–700 eV temperature and 1–3 times the solid density, as well as the hollow Al ions measured experimentally[4]. Our results show that the influence of neighboring ions and the change in ionization conditions are crucial in determining the IPD in WDP. Compared with existing experimental measurements and other analytic or more sophisticated models, our model can hold the characteristics of plasma in a wider range of plasma conditions and can be applied to ions in highly excited states, such as hollow ions. With lower computational costs and a wider application range, our work provides a promising tool for treating IPD in WDP, which will facilitate future studies of radiation and particle transport properties.

**Theory and calculation method**

In a dense plasma environment characterized by electron temperature $T_e$ and average electron density $n_0$, electrons around ions experience a screening potential induced by the target ion, plasma electrons and other ions. In this work, the contributions from neighboring ions to the screening potential are considered. In the plasma environment, neighboring ions can be populated at different ionization degrees. In the present work, to include the average influence of neighboring ions and simplify, neighboring ions are approximated in terms of the average ionization degree. In the mean field approximation and considering the contribution from neighboring ions, the screening potential can be written as

$$V_{scr}(\boldsymbol{r}) = -\frac{Z}{r} + \int_0^{r_0} \frac{n_b + \delta n(\boldsymbol{r}')}{|\boldsymbol{r}-\boldsymbol{r}'|} d\boldsymbol{r}' - \frac{Q_A}{|\boldsymbol{R}-\boldsymbol{r}|} \quad \text{for } |\boldsymbol{r}| < r_0, \quad (1)$$

Here, $Z$ is the nuclear charge number, $n_b$ is the density distribution of bound electrons of the target ion, and $\delta n(\boldsymbol{r}) = n(\boldsymbol{r}) - n_0$ is the plasma electron density fluctuation induced by the presence of ions. The plasma electron density $n(\boldsymbol{r})$ is related to the screening potential in Eq. (1). The average distance between the target ion and neighboring ions can be determined by the density of ions as $R = 2\left(\frac{3}{4\pi n_i}\right)^{1/3}$, where $n_i$ is the density of ions and where $r_0$ is the distance where $V_{scr}$ has the maximum value between the target and neighboring ions, which can be viewed as the edge of the target ion. The third term in Eq. (1) is the direct contribution from neighboring ions, and $Q_A$ is the effective charge of the neighboring ions, which can be calculated as $Q_A = Z - \int_0^{R-r_0} [n_b^A(\boldsymbol{r}') + \delta n^A(\boldsymbol{r}')] d\boldsymbol{r}'$. $n_b^A$ and $\delta n^A$ are the density of bound electrons in neighboring ion and the density fluctuation induced by the presence of neighboring ions, respectively. In a dense plasma environment, there will be several neighboring ions on average in the sphere with radius $R = 2\left(\frac{3}{4\pi n_i}\right)^{1/3}$; for simplicity, the screening potentials are approximated as spherically symmetrical as $\boldsymbol{r} \to r$ and $|\boldsymbol{R} - \boldsymbol{r}| \to |R - r|$ in the present work.

When the plasma environment reaches equilibrium, the plasma electron density $n_{FD}(r)$ can be written as

$$n_{FD}(r) = \frac{1}{\pi^2} \int_{\varepsilon_0}^{\infty} f_{FD}(\varepsilon, r) \varepsilon^{1/2} d\varepsilon, \quad (2)$$

where $f_{FD} = 1/[\exp[(\varepsilon + V_{scr} - \mu)/T_e] + 1]$ is the Fermi–Dirac distribution, $\varepsilon$ is the kinetic energy of electrons, $\varepsilon_0$ is the lower limit of $\varepsilon$, and $\mu$ is the chemical potential, which can be determined by the average density of electrons $n_0$.

In most of the present plasma models, only electrons with total energy $E = \varepsilon + V_{scr}(r) > 0$ are considered. However, in a dense plasma environment, plasma electrons can temporarily recombine with ions and distribute into negative-energy states with $E < 0$ through the TBR, DR and other inelastic collision processes between electrons and target ions. These temporarily recombined electrons also contribute to the screening potential in Eq. (2). Therefore, in the present work, electrons populated in the outermost shell are considered in the screening potential, and the orbital energy of the outermost shell is chosen as the lower limit of the total energy as $E_{min} = \varepsilon_0 + V_{scr}$ in Eq. (2).

On the other hand, as shown in Figure 1, the ionization conditions change in a dense plasma environment. As presented in Figure 1 (a), in an isolated situation, electrons being excited into states with $E > 0$ can be considered ionized. In a dense plasma environment, which is influenced by free electrons and neighboring ions, the screening potential is lower than that of isolated ions. As presented in Figure 1 (b), in previous models, the sphere surrounding an ion can be defined by the Wigner–Seitz radius $r_s = \left(\frac{3}{4\pi n_i}\right)^{1/3}$; at the edge of the ion, the screening potential $V_{scr}(r_s) < 0$, where electrons are excited into states with $V_{scr}(r_s) < E < 0$, is called "inner ionization", which can be considered ionized[39]. Nevertheless, as presented in Figure 1 (c), owing to the neighboring ions, the outer orbitals of the target ion expand into continuous bands, and electrons populated in these bands can travel from the target ion into neighboring ions, which is similar to the situation in solid materials. As a result, these electrons become delocalized, even if their total energy $E < V_{scr}(r_s)$, and electrons being excited into these bands can be considered ionized. Without considering the delocalization of these electrons, the "inner ionization" may underestimate the contribution of the variation in the ionization conditions to IPD.

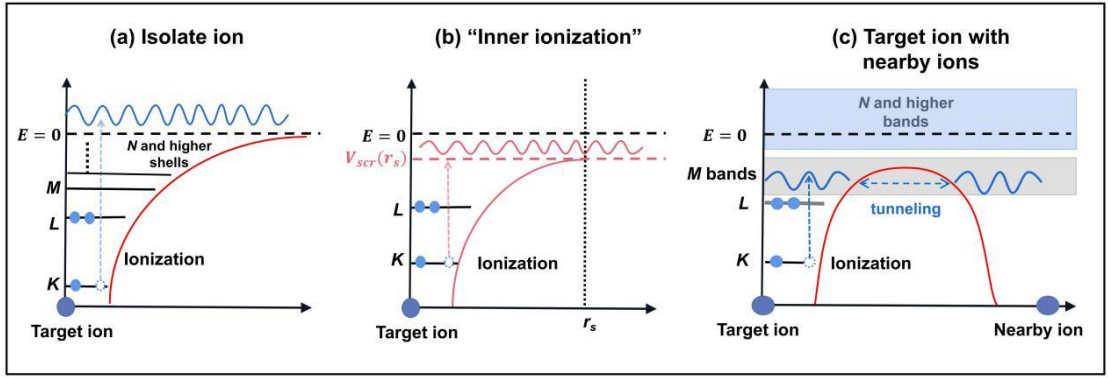

**Figure 1** (a) Schematic of ionization in an isolated ion. (b) "Inner ionization" in dense plasma. (c) Influence of neighboring ions on the target ion in WDP. The outer orbitals of the target ion expand into continuous bands, and electrons excited into these continuous energy bands become delocalized and can be considered ionized.

In the present work, the outermost shell is considered a continuous energy band, and electrons populating the outermost shell are considered ionized. The IPD of ions in ground states can be calculated as

$$\Delta I = I_0 - (E_{1snl} - E_{1s^2}), (3)$$

where $I_0$ is the ionization potential in the absence of a plasma outside, $E_{1s^2}$ is the energy of the ground state and $E_{1snl}$ is the energy of the delocalized state with $nl$ orbits in the outermost delocalized shell. In the present calculation of Al, Mg and Si plasmas with solid density and ~100 eV temperature, as presented in the experiment[4, 5], the $M$ shell is the outermost shell, and the energy of the $M$ shell should be chosen as the lower limit of the total energy as $E_{\min} = \varepsilon_0 + V_{scr}$. On the other hand, when an electron is excited into the $M$ shell, this electron is considered ionized, and the IPDs are calculated via Eq. (3).

The IPDs of different charge states of Mg, Al and Si ions are calculated by applying the present model. In detail, in the first step, the atomic orbitals (AOs) of isolated ions are prepared by using the GRASP2K code with the multiconfiguration Dirac Fock (MCDF) method[43]. Then, the density of bound electrons $n_b$ can be obtained via AOs, and the occupation numbers of the target ion are as follows:

$$n_b(r) = \sum_{i=1}^{N} N_i [P_i^2(r) + Q_i^2(r)], \quad (4)$$

where $N$ is the number of orbits, $N_i$ is the occupation number of each orbit, and $P_i(r)$ and $Q_i(r)$ are the large and small components of AOs, respectively. In the second step, the density of bound electrons is substituted into Eq. (1), the initial plasma electron density fluctuation $\delta n$ can be chosen as 0, and then, the effect potential with only the influence of the bound electrons of target ions $V_{bound}$ is obtained. In the third step, we substitute $V_{bound}$ as the trial solution into Eq. (2), and a plasma electron density fluctuation $\delta n$ related to $V_{bound}$ can be calculated from Eq. (2). By substituting $\delta n$ into Eq. (1), a new screening potential $V_{scr}$ is obtained. $V_{scr}$ is substituted into Eq. (2), and the second and third steps are repeated until $V_{scr}$ converges. Finally, IPD can be obtained by substituting the converged $V_{scr}$ into an MCDF calculation and applying Eq. (3). Notably, AOs are also influenced by the screening potential, but under present density and temperature conditions, the Debye length $D$[18], which can be used to estimate the space scale of plasma screening, has an evident influence and is still much larger than the radii of the $K$ and $L$ shells; thus, the influence of plasma screening on AOs is limited and is neglected in the present work. Compared with the other more sophisticated IPD models, the numerical costs of our model depend on the precision requirements; in the present work, it only takes seconds to compute a particular ionization degree by applying a single thread on a personal computer.

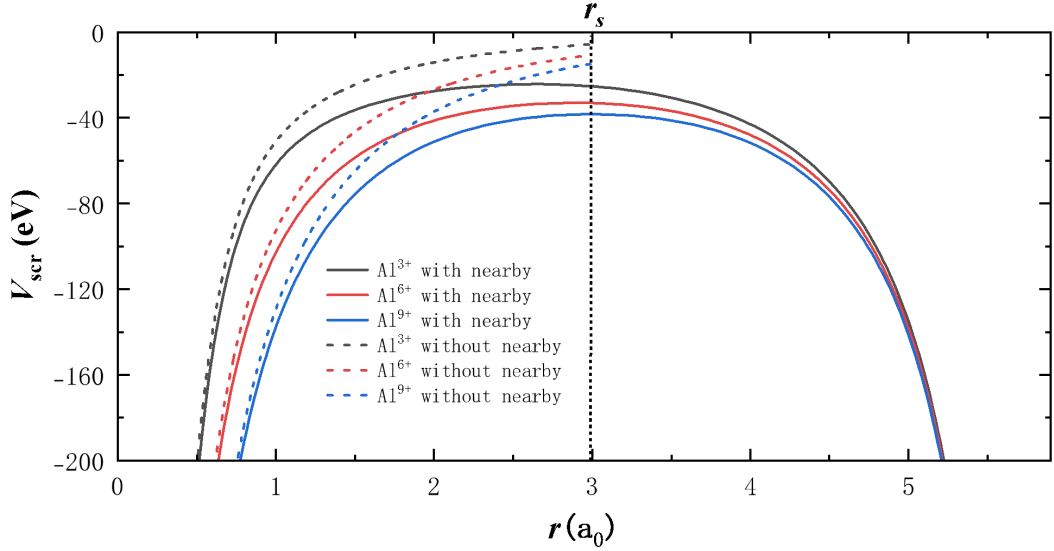

Figure 2 Effective screening potential of $Al^{3+}$, $Al^{6+}$ and $Al^{9+}$ ions with and without the influence of neighboring ions on the solid density and $T_e=100$ eV.

To validate the influence of neighboring ions, the effective screening potentials $V_{scr}$ of the $Al^{3+}$, $Al^{6+}$ and $Al^{9+}$ ions with and without the influence of neighboring ions on the solid density and $T_e=100$ eV are presented in Figure 2. Under such conditions, the average ionization degree is approximately 9, and $Al^{9+}$ ions are chosen as the neighboring ions. Figure 2 shows that neighboring ions significantly influence the screening potential, especially at the edge of the target ion. In detail, when $r = r_s$, the screening potential $V_{scr}(r_s)$ with the influence of neighboring ions is evidently greater than that without neighboring ions, and in the present work, electrons being excited into the outermost shell can be considered ionized; the total energy of these delocalized electrons is probably even lower than $V_{scr}(r_s)$. As a result, the consideration of neighboring ions influences the screening potential directly and the ionization conditions.

Under such conditions, the radial wave functions $y(r)$ of the $M$ shell are calculated. As an example, $y(r)$ of $Al^{6+}$ in the [$1s\ 2\ s^2\ 2p^3\ 3l^1$] configuration is presented in Figure 3(a). At the edge of the target ion, $y(r_0)$ of the $M$ shell orbit still deviates from 0, the electrons occupied in these orbits become delocalized, and the electrons excited into these orbits can be ionized. As presented in Figure 3(b), under such conditions, IPD includes two parts: the variation in the orbital energy of the $1s$ orbit and the change in the ionization condition, which was not adequately considered in previous work.

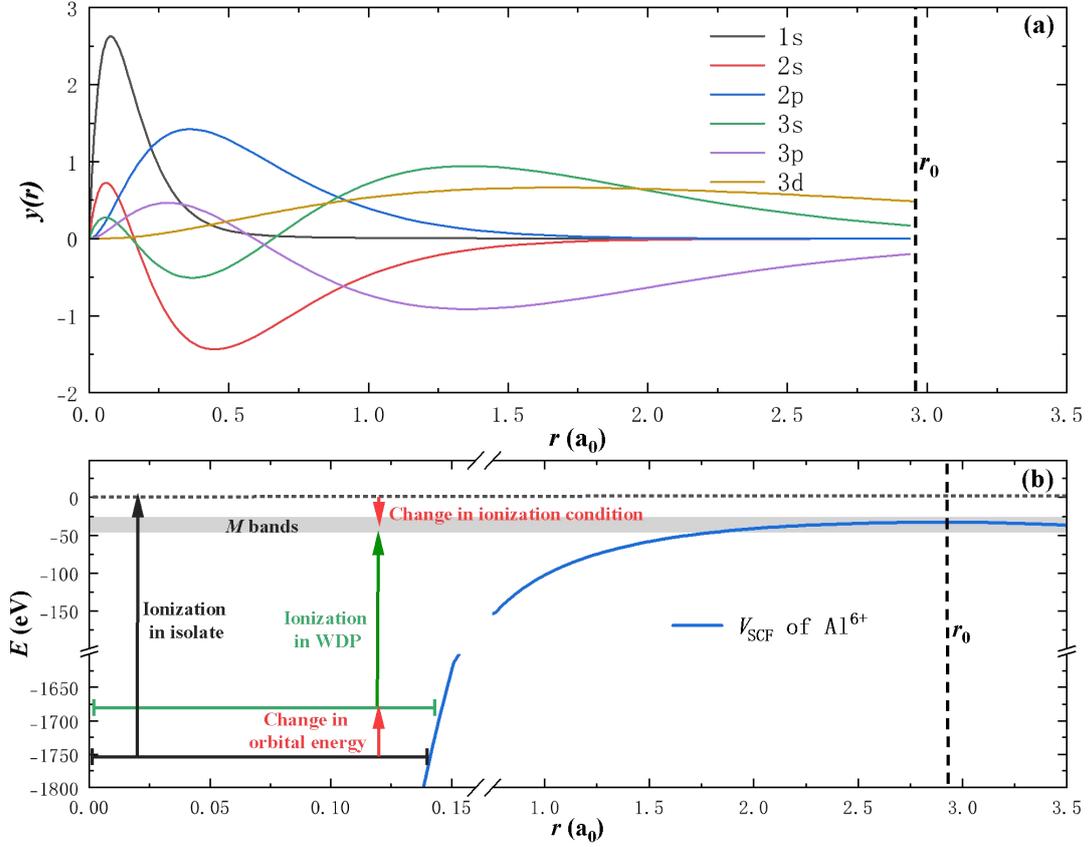

Figure 3 (a) Radial wave functions $y(r)$ of an Al configuration [$1s\ 2s^2\ 2p^3\ 3l^1$] with solid density and $T_e$=100 eV. (b) Schematic diagram of the IPD of $Al^{6+}$; under such conditions, the IPD includes two parts: the variation in the orbital energy of the 1 s orbital and the change in the ionization conditions.

## Results and Discussion

The present model is first applied to calculate the IPD in plasma conditions of LCLS experiments, in which the density of ions $n_i$ is similar to that in solid materials and the electron temperature $T_e$ ranges from 70 to 180 eV. In the LCLS experiment, the ionization of a *1 s* electron by the incident laser results in the creation of an ion with a hole in the *K*-shell, followed by a *Kα* emission to fill the hole in the *K*-shell. Therefore, the signal of the *Kα* emission is directly proportional to the intensity of the incident laser. The incident X-ray pulse reaches its peak at 80 fs[5], leading to the strongest emitted signal. Additionally, Ref. [3] demonstrated that the temperature of the plasma increases during a laser pulse, reaching approximately 70 to 100 eV at the peak of the X-ray pulse. Consequently, in this study, the IPDs of Al and Mg Si ions with $T_e$=70 and 100 eV and $n_i$ at the solid material density are calculated in the present model and compared with the experimental results, the results of which are presented in Figure 4. To validate the influence of neighboring ions, the IPDs calculated without the contribution from neighboring ions in $V_{scr}$ and using the "inner-ionization" condition are also presented. Such IPDs can be calculated as $\Delta I_0 = \langle \Psi | V_{scr} - V_{iso} | \Psi \rangle - V_{scr}(r_s)$, where $|\Psi\rangle$ is the atomic orbital function (AO) of the bound electron, $V_{iso}$ is the effective potential felt by an isolated ion, and the electron can be ionized when its energy becomes higher than $V_{scr}(r_s)$ [39]. For comparison, the IPDs calculated via the SP and EK models are also presented.

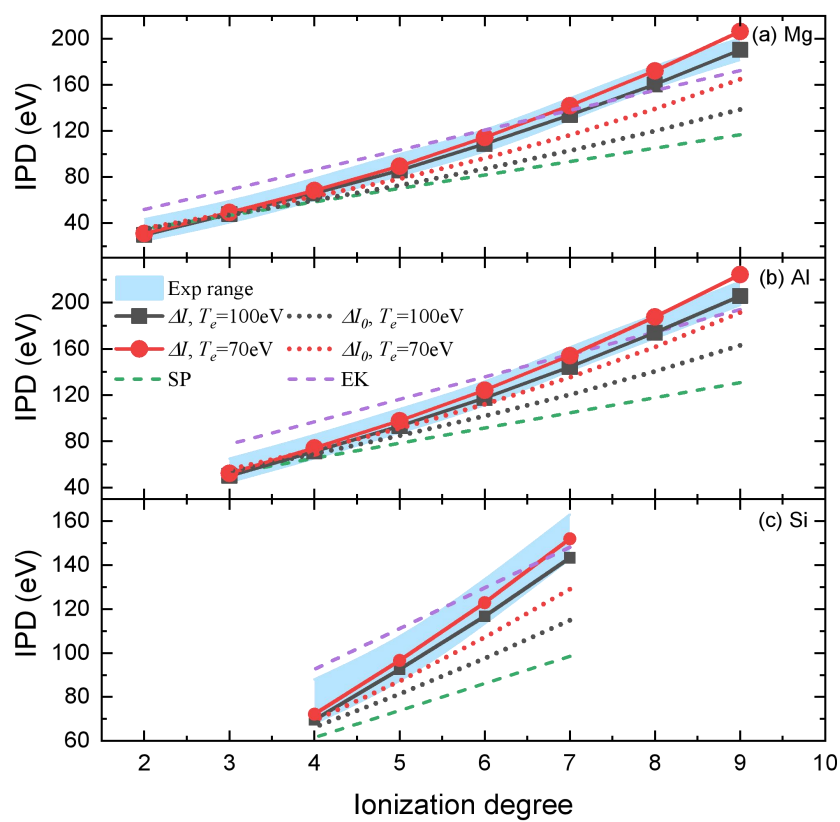

Figure 4 IPD calculated in the present model and comparison with the LCLS experiment. (a), (b), (c): IPDs of Mg, Al and Si ions with solid density and $T_e$ = 70 and 100 eV, respectively. $\Delta I$ represents the results of the present model, and $\Delta I_0$ represents the results without the influence from neighboring ions in $V_{scr}$ and uses the widely applied "inner-ionization" conditions.

As shown in Figure 4 (a-c), the IPDs calculated in the present model are in good agreement with the experimental values[4, 5]. In detail, both the IPDs with $T_e = 70$ eV and 100 eV are within the experimental error bar for Mg, Al and Si ions. For low ionization degrees such as $Mg^{2+}$, $Al^{3+}$ and $Si^{4+}$ ions, the results with $T_e = 70$ eV and 100 eV are fairly close. For higher ionization degrees, the difference between the results with $T_e = 70$ eV and 100 eV increases. These results indicate a stronger temperature dependence in the IPD of ions with higher ionization degrees. On the other hand, as presented in Figure 4, without considering the contribution from neighboring ions, the results $\Delta I_0$ are lower than the experimental range, which underestimates the IPDs under such conditions. These results indicate that the influence of neighboring ions is important and should be considered under such conditions.

Among the previous widely applied analytic models, in the plasma conditions related to the LCLS experiments, the IPR-estimated EK model seems to be in better agreement with the measurements, whereas the SP model underestimates the IPDs under such conditions. However, in higher density and temperature conditions related to the Orion experiments[6], EK models overestimate the IPD evidently, whereas the SP model has more or less better agreement with the measurements. In addition to the LCLS and Orion experiments, the IPDs of $Mg^{7+}$ ions with solid density and $T_e = 75$ eV were obtained in the latest experiment[11, 36]. Therefore, for further validation of the present model, the IPDs under plasma conditions are related to the Orion experiments and are also calculated, and the results are presented in the table. 1.

Table 1 IPDs calculated under plasma conditions related to the Orion experiment and compared with other analytic models and experimental measurements.

| Plasma conditions | Ionization degrees | EK model (eV) | SP model (eV) | Present model (eV) | Expt. (eV) |
|---|---|---|---|---|---|
| $\rho = 5.5 \text{ g/cm}^3$, $T_e = 550$ eV | $Al^{11+}$ | 326 | 190 | 177 | <224 |
| | $Al^{12+}$ | 353 | 206 | 201 | <256 |
| $\rho = 9 \text{ g/cm}^3$, $T_e = 700\ eV$ | $Al^{11+}$ | 387 | 224 | 260 | >224 |
| | $Al^{12+}$ | 419 | 242 | 278 | >256 |
| $n_e = 3\text{e}23/\text{cm}^3$, $T_e = 75$ eV | Mg$^{7+}$ | 130 | 89 | 138 | 132 |

As presented in Table 1, under conditions where $n_0 = 3 \times 10^{23}/\text{cm}^3$, and $T_e = 75$ eV, which are close to the conditions of the LCLS experiments[4, 5], our results are close to those of the EK model, which are in good agreement with the measurements, and the SP model evidently underestimates the IPDs. However, at higher temperatures and densities, our model approaches the SP model. In detail, when the density of the plasma is $\rho = 5.5 \text{ g/cm}^3$ and $T_e = 550$ eV, the Orion experiment results in transitions between the *3p* and *1s* orbits in the Al$^{11+}$ and Al$^{12+}$ ions, and the *M* shell still exists in such cases. Therefore, in the present calculation, the orbital energy of the *N* shell is chosen as the lower limit of the total energy $E_{min}$. In such cases, our results are close to those of the SP models, which are smaller than the ionization potential of the *K* shell, and the *M* shell still exists, as the experimental results show. The EK model significantly overestimates the IPDs, which are even higher than the ionization potential of the *M* shell. As a result, the *M* shell merges into a continuum, which is inconsistent with the experimental results. When $\rho = 9 \text{ g/cm}^3$ and $T_e = 700$ eV, the experiment shows that both the transitions between the *3p* and *1 s* orbitals in Al$^{11+}$ and Al$^{12+}$ vanish, which demonstrates that the IPDs of Al$^{12+}$ and Al$^{11+}$ have exceeded the ionization potential of the *M* shell, i.e., 220 and 256 eV, respectively. On the other hand, in such a case, the IPD of Al$^{12+}$ calculated in the SP model is still lower than the ionization potential, which indicates a slight underestimation of the IPD under such conditions. In the present calculation, because the *M* shell merges into a continuum, the orbital energy of the *M* shell is chosen as the lower limit of the total energy $E_{min}$. As presented in Table 1, under such conditions, the IPDs calculated in the present model exceeded the ionization potential for both the Al$^{11+}$ and the Al$^{12+}$ ions, which is in good agreement with the experimental results. From Figure 4 and Table 1, compared with the traditional analytic SP and EK modes, our model can be applied under wider plasma conditions and has better agreement with existing experiments[4-6].

In LCLS experiments, the *Kα* lines of hollow ions with one *K*-shell electron ionized are also observed; these lines provide our chance to obtain the IPD of hollow ions[4, 5, 44]. Compared with ions in ground states, these hollow ions with one *1* s electron have a very short lifetime because of the fast auger and radiation processes. As a result, the electron distribution around these ions may deviate from LTE conditions. The IPD of such highly excited hollow ions cannot be calculated by the widely applied SP and EK models, as they are not state-resolution models, and is also difficult for DFT, as DFT is focused mainly on ions in ground states.

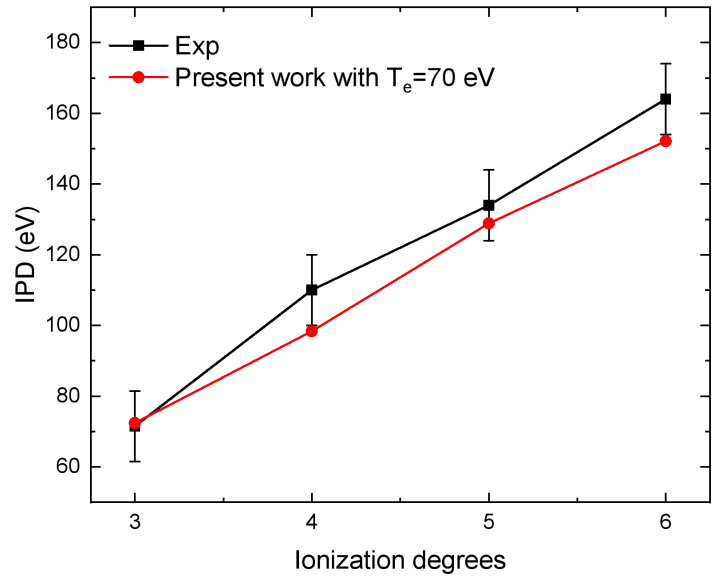

Figure 5 IPD of hollow *1 s* Al ions with solid density and $T_e = 70$ eV in the present model compared with the experimental results.

The IPDs of the hollow 1 s Al ions are calculated with the solid density and $T_e = 70$ eV in our model and compared with the experimental values[4, 5]. The results are presented in Figure 5. The uncertainty of the measurements is 10 eV, which is estimated by the IPD of ions in ground states, but owing to the much weaker signal, the uncertainty may be underestimated. The results of our model are in good agreement with the measurements. The slight discrepancy may originate from the underestimation of the uncertainty in experiments or from the non-LTE distribution of electrons around these hollow ions, such as the selection effect of the pump laser[42].

These results indicate that, compared with widely applied SP and EK models or more sophisticated DFT methods, our model is a state-resolution model and can be applied in the calculation of highly excited ions.

Finally, as we discuss in Figure 4, in our model, the IPD of ions with higher ionization degrees has a stronger temperature dependence. Therefore, the IPDs of $Al^{3+}$, $Al^{6+}$ and $Al^{9+}$ with respect to the solid density and different temperatures are calculated. For comparison, the results without considering the influence of neighboring ions $\Delta I_0$ and the results from the SP and EK models are also calculated. The results are presented in Figure 6. For ions with lower ionization degrees, our results are close to those of the SP models, whereas for ions with higher ionization degrees, our results are between those of the SP and EK models. In the lower temperature area, our model has an evident temperature dependence, especially for ions with higher ionization degrees. However, with increasing temperature, the temperature dependence becomes weak, and the variation tendency is similar to that of the SP models, which have a very weak temperature dependence. As discussed in our previous work, the temperature dependence is from the consideration of the temporarily recombined electrons[42]. In the high-temperature area, the distribution of the temporarily recombined electrons becomes negligible, and the temperature dependence becomes weak. Compared with the results of $\Delta I_0$, the temperature dependence of the present model is slightly weaker. These results indicate possible competition between temporarily recombined electrons and the influence of neighboring ions. On the other hand, the EK model is related only to the density of electrons and ions; as the temperature increases and the density of ions remains constant, the density of electrons increases; as a result, the IPD calculated by the EK model even increases with increasing temperature in such cases.

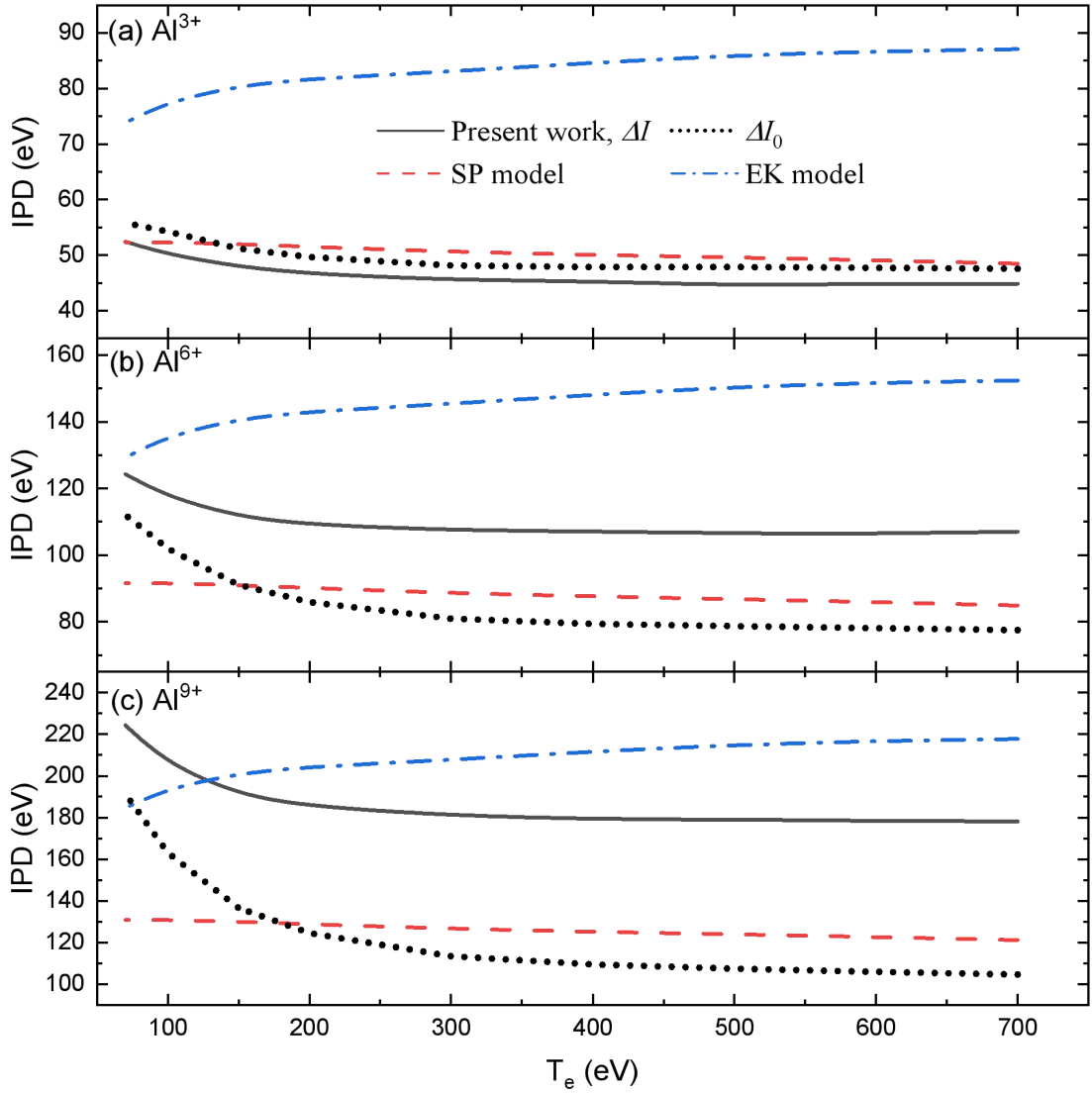

Figure 6 IPD of $Al^{3+}$, $Al^{6+}$ and $Al^{9+}$ ions as a function of temperature (from 75 to 700 eV) with respect to solid density and comparison with the results from the SP and EK models.

**Conclusion**

In conclusion, under WDP conditions, the influence of neighboring ions becomes nonnegligible. The contribution from neighboring ions to IPD can be divided into two parts. First, neighboring ions directly influence the screening potential around the target ion, which then changes the ionization potential. Second, the outer orbitals of the target ion expand into continuous bands because of the existence of neighboring ions, and electrons in these continuous bands can travel from the target ion into neighboring ions and become delocalized. As a result, even if their total energy $E <$

0, electrons excited into these continuous bands can be considered ionized, the ionization conditions change compared with those in isolated situations. By including the direct contribution from neighboring ions in the screening potential of the target ion and considering the variation in ionization conditions due to neighboring ions, we refine our atomic-state-dependent screening model, which can be applied under WDP conditions. The developed model is successfully validated by effectively reproducing the IPDs in recent warm and dense plasma experiments under different temperature and density conditions. Our results indicate that the contribution from neighboring ions is important and should be accounted for when accurately modeling the IPD in WDP. Compared with existing experimental measurements and other analytic or more sophisticated models, our model can hold the characteristics of plasma in a wider range of plasma conditions and can be applied to ions in highly excited states, such as hollow ions. With lower computational costs and a wider application range, our work provides a promising tool for treating IPD in WDP, which will facilitate future studies of radiation and particle transport properties.

**Acknowledgements**

This work is supported by the National Natural Science Foundation of China under Grants No. 12204057 and No. 62165007, the Yunnan Applied Basic Research Projects under Grant No. 202401CF070090, and the Foundation of the National Key Laboratory of Computational Physics. We acknowledge the computational support provided by the Institute for Applied Physics and Computational Mathematics.

**Author contributions**

**Chensheng Wu**: Developed the theoretical framework, implemented computational codes, and write the original manuscript. **Jiao Sun**: Performed numerical simulations, conducted data analysis, and co-wrote the original manuscript. Chunhua Zeng and Qinghe Song: Participated in the conceptualization and mathematical formulation of the theoretical model. **Xiang Gao and Jun Yan**: Provided critical theoretical guidance, validated methodological rigor, and oversaw manuscript revisions.

**Data availability statement**

The datasets generated and analyzed during this study are available from the corresponding author upon reasonable request.